\documentclass[preprint,nofootinbib,prd,aps]{revtex4}
\usepackage{graphicx,amstext,amssymb,amsmath,color,ulem,hyperref}
\begin{document}

\title{Particle   momentum spectra, correlations, and maximum entropy principle in high-multiplicity collision events  }

\author{S.V. Akkelin$^{1}$}

\affiliation{$^1$Bogolyubov Institute for Theoretical Physics,
Metrolohichna  14b, 03143 Kyiv,  Ukraine}

\begin{abstract} 

 In this paper, we utilize the maximum entropy prescription to determine a quantum state of a small collision system at the kinetic freeze-out.  We derive expressions for multiplicity-selected particle momentum spectra and correlation functions by applying a fixed particle number constraint to this state.  The results of our analysis can be useful for interpreting the multiplicity dependence of the particle momentum spectra and correlations in high-multiplicity  $pp$ collision events at a fixed LHC energy.

\end{abstract}


 \maketitle

\section{Introduction}

With the advent of the Large Hadron Collider (LHC),  dividing a whole set of proton-proton ($pp$) collision events at a fixed energy into subsets with
fixed charged-particle multiplicities became possible. It is widely accepted that an increase in charged-particle
multiplicity in $pp$ collision events at a fixed energy is conditioned by the rise in the initial geometrical size of
the reaction zone or/and an increase in the initial energy density (for a review, see, e.g., Ref. \cite{initial}). The former reaches its maximum for central collisions, where optimal overlapping of colliding protons is reached. 
The latter can fluctuate over the mean value corresponding to a given impact parameter, that is, at a fixed centrality of a collision. Then it is expected that, for central collisions, where the maximal initial size of the reaction zone is reached, an increase in charged-particle multiplicities above a mean value at the given centrality might be conditioned by an increase above a mean value in the initial energy densities. 
High-multiplicity events in such collisions appear to have features in common with the hydrodynamical picture of relativistic heavy ion collisions; for reviews, see, e.g., Refs. \cite{S,W}.

Suppose a hydrodynamic description is
applicable for $pp$ collisions. In that case, an increase in the energy density results in an increase in pressure gradients, and the latter increases the intensity of a collective expansion. As is well known, an increase in collective expansion reveals itself, in particular, in an increase in the mean transverse momenta of particles. The latter have been measured
and indeed increase with multiplicity \cite{CMS-spectra,Alice-spectra}, while in the very high-multiplicity range this increase is not significant \cite{CMS-spectra}.\footnote{See also Ref. \cite{Biro} where it was demonstrated that in the low transverse momentum region, $p_{T}<0.6$ GeV,  the mean transverse momenta do not increase 
 with multiplicity.}  Also, the multiplicity dependence of the  two-particle  Bose-Einstein 
 momentum correlation radii\footnote{These radii are the result of the fit of the correlation function
 defined as a ratio of the two-particle spectra to the product
of the single-particle ones. For a review of the correlation femtoscopy method, see e.g.
Refs. \cite{HBT-1,HBT-2,HBT-3}.} at the LHC has been studied at
a fixed energy of collisions.  
One notable feature of these measurements is that the effective system’s volume,
when extracted from the  correlation radius parameters, appears to scale nearly
linearly with charged-particle multiplicity  \cite{Atlas,CMS,Alice}, except for low multiplicities 
where initial state effects dominate, and quite unexpectedly, except for very high multiplicities
where, at variance with the expected behavior for emission from hydrodynamically
expanding systems (see, e.g., Ref. \cite{Sinyukov-1}), one sees systematic deviations
from such a scaling behavior. That is, the striking feature of the data is that the correlation
radius parameters become approximately independent of the particle multiplicity in the 
limit of very high multiplicities \cite{Atlas,CMS}. The question naturally arises:
How can this controversy with hydrodynamics be resolved?

First, note that the initial state in $pp$ collisions is a quantum state, and a quantum state cannot be considered only to represent the statistical properties of an ensemble of similarly prepared systems.
Generally speaking, a statistical ensemble of events
(with corresponding probabilities) appears only
when measurements are performed in a large number of experiments made under identical conditions.
Classical-statistical approximation of the initial state in small collision systems can be valid for 
sufficiently high occupation numbers, but breaks down for a more dilute
state where quantum effects are important, which is reached rather rapidly
via expansion. This leads to effects of the quantum uncertainty and, therefore, could result in interference
between alternatives started from the different eigenstates of the density projection operators.
Such an interference prevents attributing probabilities to evolutions started from the initial
density fluctuations and, therefore, might invalidate the applicability of 
hydrodynamics associated with different energy density eigenstates at a fixed centrality 
of a collision.\footnote{Validity of the hydrodynamic 
description of small quantum
systems was questioned many years ago based on the quantum uncertainty principle in Ref. \cite{Bl}. 
For further discussions, see, e.g., Refs.  \cite{Halliwell,Gell-Mann} and references therein.}  

At the same time, a hydrodynamic description of the evolution may still be applied for the mean density
associated with the expectation value of the energy-momentum tensor in a quantum state corresponding to, say, the most central $pp$ collisions. To proceed with such a description, the
true state of a system can be approximated by a state that is characterized by the knowledge
of the expectation values of only some observables. This can be done by using Zubarev's formalism of the nonequilibrium statistical
operator \cite{Zubarev-1a,Zubarev-1b,Zubarev-1c,Zubarev-1d}; for
modern developments, see Refs. \cite{Zubarev-2,Zubarev-3,Zubarev-4} and references therein.
 In such an approach, a statistical ensemble of multiplicity
fluctuations, corresponding to some fixed centrality of a collision, can appear at some later stage of evolution, when a classical-statistical
description of these fluctuations
becomes possible. Once this is taken into account,  the discrepancy with hydrodynamics should disappear.
  
This paper is structured as follows. 
In Sec. \ref{decoherence}, we focus on the distinction between quantum and classical multiplicity fluctuations.
In Sec. \ref{maximum}, we apply the maximum entropy principle to estimate a quantum state at the kinetic freeze-out. In Sec.  \ref{HBT}, we derive the expressions
 for multiplicity-selected particle momentum spectra and correlation functions.
Our conclusions are given in Sec. \ref{Concl}.

\section{Quantum interference and decoherence   of  multiplicity fluctuations  }
\label{decoherence}

It is natural to assume  that different values of the impact parameter  (i.e., different centralities of a collision)
can be associated with mutually orthogonal quantum states of a small collision system. Then,  a $pp$ collision can be presented as a
statistical ensemble of collisions with various impact parameters. Among these states, 
there is a state that corresponds to the maximal initial geometric size of the interaction region. 
This implies that multiplicity fluctuations arise in $pp$ collisions at a fixed energy of collisions  because of an incoherent superposition 
of collisions with  different values of the impact parameter, and because of the quantum fluctuations of  
multiplicity at a fixed impact parameter.
The question is whether a statistical ensemble of particle multiplicities corresponding to a fixed centrality of a collision may be introduced before the measurement is done.

To answer this question, recall that one cannot prescribe probabilities to not measured histories 
if there is a quantum interference between them, because then the sum rules of probability theory are not 
satisfied: In the quantum theory, probabilities are squares of amplitudes. The double-slit experiment provides
an elementary example: the probability of arriving at a point on the screen is not the sum of probabilities
to go through the alternative slits. It happens because in the course of evolution, an eigenvalue
of the position operator does not evolve with time in some other one
but into a superposition (so-called spreading of a wave pocket), and because two wave pockets overlap behind 
the slits. If the screen is situated very close to the slits, then one can imagine a situation
where the wave pockets do not overlap. Then probabilities can be assigned for histories (paths) before 
actual measurement (on the screen) has been done.  
Hence, one can conclude that probabilities can be assigned for alternative histories of a closed system
if there is no interference between them. It should be achieved, for a given Hamiltonian, by an appropriate initial condition.   

We are interested here in histories consisting of projections onto particle numbers of free (noninteracting) particles.
The corresponding  Hermitian projection operator ${\cal P}_{N}$ in the  Schr\"odinger picture is 
\begin{eqnarray}
{\cal P}_{N} =\int  d^{3}p_{1}... d^{3}p_{N} |p_{1},...,p_{N}\rangle
\langle p_{1},...,p_{N} | ,  \label{1} 
\end{eqnarray}
where $|p_{1},...,p_{N}\rangle$ is defined as 
\begin{eqnarray}
|p_{1},...,p_{N}\rangle =\frac{1}{\sqrt{N!}} a^{\dag}(\mathbf{p}_{1})...a^{\dag}(\mathbf{p}_{N})| 0 \rangle .
\label{2}
\end{eqnarray}
Here $a^{\dag}(\textbf{p})$ and $a(\textbf{p})$
are creation and annihilation operators, respectively, which
satisfy the following canonical commutation relations\footnote{To focus on the basic issues, we consider  a neutral scalar field hereafter.}:
\begin{eqnarray}
[a(\mathbf{p}), a^{\dag}(\mathbf{p}')] =
\delta^{(3)}(\mathbf{p}-\mathbf{p}'), \label{3}
\end{eqnarray}
and 
$[a(\textbf{p}), a(\textbf{p}')]=[a^{\dag}(\textbf{p}), a^{\dag}(\textbf{p}')]=0$.
One can see that $\hat{N}{\cal P}_{N}=N{\cal P}_{N}$ where
\begin{eqnarray}
\hat{N}=\int d^{3}p a^{\dag}(\textbf{p})a(\textbf{p}) \label{3.1}
\end{eqnarray}
 is the particle number operator.

The operators ${\cal P}_{N}$ are exhaustive, 
\begin{eqnarray}
\sum_{N=0}^{\infty}{\cal P}_{N}={\cal I},
 \label{4}
\end{eqnarray}
and mutually exclusive,
\begin{eqnarray}
{\cal P}_{N_{i}}{\cal P}_{N_{j}}=\delta_{ij}{\cal P}_{N_{i}}.
 \label{5}
\end{eqnarray}
In the Heisenberg picture, the operator  ${\cal P}_{N}$ evolves with time according to 
\begin{eqnarray}
{\cal P}_{N}(t)=e^{iHt}{\cal P}_{N}(0)e^{-iHt},
 \label{6}
\end{eqnarray}
where ${\cal P}_{N}(0)$ coincides with the projection operator in the Schr\"odinger picture,
${\cal P}_{N}(0)={\cal P}_{N}$.
One can see that operators ${\cal P}_{N}(t)$ are exhaustive and mutually exclusive for each time point $t$.
For noninteracting fields, the particle number operator, $\hat{N}=\int d^{3}p a^{\dag}(\textbf{p})a(\textbf{p})$,
is conserved and commutes with the  Hamiltonian, i.e., $[\hat{N},H]=0$. Then  the projection operator ${\cal P}_{N}(0)$ commutes with  the Hamiltonian, and 
\begin{eqnarray}
{\cal P}_{N}(t)=e^{iHt}{\cal P}_{N}(0)e^{-iHt}={\cal P}_{N}(0).
 \label{7}
\end{eqnarray}
In general, a particle number is not a conserved quantity for interacting fields.
Then  $[\hat{N},H] \neq 0$, 
${\cal P}_{N}(t)\neq {\cal P}_{N}(0)$  and ${\cal P}_{N}(t)$ contains projections into  various particle numbers. 

A quantitative measure  of the interference of different histories is provided by the 
decoherence functional, see e.g. Ref. \cite{Gell-Mann}.
Consider the two-time decoherence functional $D(N_{2},N_{1};N'_{1},N_{2})$, which is associated with 
two different paths to the same final state. In the Schr\"odinger picture, 
\begin{eqnarray}
D(N_{2},N_{1};N'_{1},N_{2})=\mathrm{Tr}[{\cal P}_{N_{2}}e^{-iH(t_{2}-t_{1})}{\cal P}_{N_{1}}e^{-iHt_{1}}\rho (0) e^{iHt_{1}}{\cal P}_{N'_{1}}e^{iH(t_{2}-t_{1})}{\cal P}_{N_{2}}].
 \label{7.1}
\end{eqnarray}
 Here $\rho (0)$ denotes the initial quantum state (density matrix).
Equivalently, in the Heisenberg picture,
\begin{eqnarray}
D(N_{2},N_{1};N'_{1},N_{2})=\mathrm{Tr}[{\cal P}_{N_{2}}(t_{2}){\cal P}_{N_{1}}(t_{1})\rho {\cal P}_{N'_{1}}(t_{1}){\cal P}_{N_{2}}(t_{2})]= \nonumber \\
\mathrm{Tr}[{\cal P}_{N'_{1}}(t_{1}){\cal P}_{N_{2}}(t_{2}){\cal P}_{N_{1}}(t_{1})\rho ].
 \label{8}
\end{eqnarray}
Here $\rho=\rho (0)$, and   
we used the cyclic invariance of the trace.
It is clear from Eq. (\ref{8}) that  if   $[\hat{N},H] \neq 0$, then one can expect that
${\cal P}_{N'_{1}}(t_{1}){\cal P}_{N_{2}}(t_{2}){\cal P}_{N_{1}}(t_{1}) \neq 0$ for  $N_{1}\neq N'_{1}$.
Hence,  for interacting fields, the decoherence functional (\ref{8}) can contain off-diagonal terms responsible for interference effects.
These off-diagonal terms are a measure of the degree of interference between pairs 
of alternatives as a result of spreading and overlap. When this interference is absent, the diagonal terms 
provide the probabilities. 
The situation is similar to the one in the double-slit 
experiment, see, e.g.,  Ref. \cite{Hartle-1}. 

Exact decoherence (i.e., absence of the interference) in an isolated quantum system takes place for exactly conserved 
quantities that commute with the Hamiltonian \cite{Hartle-2}. If this is not the case, then 
only approximate decoherence can occur when the interference is sufficiently low.\footnote{Such an approximate decoherence can take place, for example,  
if evolution of the projectors, see Eq. (\ref{6}),  can be approximated in the decoherence functional
as evolution of the number of particles, ${\cal P}_{N}(t)\approx {\cal P}_{N(t)}$.} Thus,  for systems created in $pp$ collisions, a statistical ensemble of particle multiplicities can be introduced at the system's decoupling, when mean densities become sufficiently low 
to approximate the evolution by a free streaming.
Then the Hamiltonian can be approximated by the noninteracting one,  and the particle number is approximately a conserved quantity.
It suggests that one can assign a probability to produce 
$N$ particles to the  diagonal element of the decoherence functional,  
\begin{eqnarray}
p(N)=D(N;N)=\mathrm{Tr}[{\cal P}_{N}\rho {\cal P}_{N}].
 \label{9}
\end{eqnarray}
Here $\rho$ should be defined at some hypersurface where 
switching off the interactions does not change the final particle momentum spectra and multiplicities. One can also define at this hypersurface the normalized state  conditional on $N$,
\begin{eqnarray}
\rho_{N} =\frac{{\cal P}_{N}\rho {\cal P}_{N}}{\mathrm{Tr}[{\cal P}_{N}\rho {\cal P}_{N}]}=\frac{{\cal P}_{N}\rho {\cal P}_{N}}{p(N)}.
 \label{10}
\end{eqnarray}

\section{Quantum maximum entropy  state at kinetic  freeze-out  } 
\label{maximum}

Our next step is the definition of $\rho$. Before we begin, we briefly note that the centrality dependence of the multiplicity reflects its
correlation with the initial geometric size of the reaction zone. Therefore, one can assume 
that the dominant
contribution to (very)  high-multiplicity collision events (which interest us here) is expected to occur from central collisions with maximal geometric size of the initial reaction zone. This means that, 
for very high-multiplicity 
events,  the $\rho$   should, in some approximation, be associated with such central collisions.
Of course, to get $\rho$ after the system's decoupling by calculating full quantum evolution is an unrealistic task. 
Instead, one can use a reduced description. Namely, one can take into account that,
at a later kinetic (posthydrodynamical) stage of evolution 
of an expanding system, its state at a fixed centrality of collisions can be described by the
 one-particle distribution function,\footnote{For simplicity, we restrict our consideration to
the situations where the one-particle distribution function is the only independent parameter. If this is not 
the case, the additional many-particle distribution functions should be considered. }   whose initial conditions are matched with hydrodynamics. 
The quantum analogue of the classical one-particle distribution function is the one-particle Wigner function,  which
 is defined as the expectation value of the corresponding  Wigner operator,  see, e.g., Ref. \cite{Groot}.  
 The  Wigner function is similar to the classical phase-space density, but the analogy has its limitations: 
 it is not positive definite, and the four-momentum is not confined to the mass shell \cite{Groot}. 

The freeze-out hypersurface for collisions with a fixed centrality is the space-time surface at which the mean particle densities, corresponding to such a centrality,  become so low that beyond this hypersurface interactions between particles cease, and particle momentum spectra are ``frozen.''  It provides that  the  
 Wigner function after kinetic freeze-out can be approximated by the  Wigner function of a free scalar field, and its evolution can be approximated by the free streaming (see Refs. \cite{Groot,Tinti}).
The average  one-particle  momentum spectrum at a fixed centrality\footnote{For the sake of simplicity, we suppress the dependence on the centrality of collisions.}   is given by
\begin{eqnarray}
p_{0}\frac{d^{3}\langle N \rangle}{d^{3}p} =   \int_{\sigma}  d \sigma_{\mu}p^{\mu}f(x,p),
\label{10.1} 
\end{eqnarray}
where $\sigma $ is a three-dimensional
freeze-out spacelike hypersurface,  
  $d\sigma_{\mu}\equiv d\sigma n_{\mu}$,  $n_{\mu}$ is a  timelike normal vector,
 $\langle N \rangle$  is a mean multiplicity in, say, 
most central collisions, and $f(x,p)$ is the corresponding one-particle Wigner function of a noninteracting scalar field \cite{Groot}.  
Given the known values of the one-particle Wigner function $f(x,p)$\footnote{ It can take on negative values, see, e.g., Ref. \cite{Groot}.} and assuming that other details are irrelevant at the freeze-out
hypersurface, the state of a system at this hypersurface can
be approximated by the statistical operator $\rho$
 based on
the principle of maximum entropy \cite{Jaynes}. Such a statistical operator 
maximizes the von Neumann entropy, $S=-\mathrm{Tr}[\rho \ln \rho]$,  subject to
the constraints
\begin{eqnarray}
f(x,p) = \mathrm{Tr} [\hat{f}(x,p)\rho]
 \label{11}
\end{eqnarray}
at a fixed centrality of collisions, where $\hat{f}(x,p)$ is the one-particle Wigner operator of a noninteracting neutral scalar field, and $f(x,p)$ is the one-particle Wigner function associated with the average one-particle momentum spectra at a fixed centrality, see Eq. (\ref{10.1}). Then, the
statistical operator $\rho$ is  given by that of a generalized Gibbs state
\begin{eqnarray}
\rho = \frac{1}{Z}\exp \left (- \int_{\sigma} d\sigma'_{\mu}\int d^{3}p'\frac{ p'^{\mu}}{p_{0}'} \lambda (x',p')\hat{f}(x',p')\right ),
 \label{12}
\end{eqnarray}
where $\lambda (x',p')$ is a real function  defined by the condition (\ref{11}), and $Z$ is a normalizing factor,
ensuring that
$\mathrm{Tr} [\rho] =1$. In what follows, we assume that    the operator $\int_{\sigma} d\sigma'_{\mu}\int d^{3}p'\frac{ p'^{\mu}}{p_{0}'} \lambda (x',p')\hat{f}(x',p')$  
is bounded from below on physical states. 

Inspired by Ref. \cite{Groot}, we present here a simple derivation of the one-particle Wigner operator for a free neutral scalar field. First, one can take into 
account that, because of (\ref{10.1}) and (\ref{11}),  
\begin{eqnarray}
p_{0}\frac{d^{3}\langle N \rangle}{d^{3}p}  =   \int_{\sigma}  d \sigma_{\mu}p^{\mu}\langle \hat{f}(x,p) \rangle,
\label{13} 
\end{eqnarray}
where $\langle ... \rangle = \mathrm{Tr} [...\rho] $.
On the other hand, by definition
$\langle N \rangle = \int d^{3}p \langle a^{\dag}(\textbf{p}) a(\textbf{p}) \rangle$, and therefore we also have 
\begin{eqnarray}
p_{0}\frac{d^{3}\langle N \rangle}{d^{3}p} = p_{0}\langle a^{\dag}(\textbf{p}) a(\textbf{p}) \rangle . 
\label{14} 
\end{eqnarray}
Combining Eqs. (\ref{13}) and (\ref{14}) one can suggest  that  the one-particle operator-valued function,  $\hat{f}(x,p)$, satisfies the equality
\begin{eqnarray}
 a^{\dag}(\textbf{p}_{1}) a(\textbf{p}_{2}) = \int_{\sigma}  d \sigma_{\mu}\frac{p^{\mu}}{p^{0}}e^{i\Delta p x}\hat{f}(x,p).
\label{15} 
\end{eqnarray}
Here  $\Delta p  =p_{2}-p_{1}$, and $p=(p_{1}+p_{2})/2$. One can see that  it is the case if 
\begin{eqnarray}
\hat{f}(x,p) =\frac{1}{(2\pi)^{3}} \int d^{4}u e^{-iux} \delta (pu) p^{0}a^{\dag}\left(\mathbf{p}-\frac{\mathbf{u}}{2}\right) a\left(\mathbf{p}+\frac{\mathbf{u}}{2}\right).
 \label{16}
\end{eqnarray}
  Indeed, taking into account that $p^{\mu}\partial_{\mu}\hat{f}(x,p)=0$ and  using the Gauss theorem, we get 
\begin{eqnarray}
\int_{\sigma}  d \sigma_{\mu}p^{\mu}e^{i\Delta p x}\hat{f}(x,p) = \int_{t}  d^{3}r p^{0}e^{i\Delta p x}\hat{f}(x,p)= \nonumber \\ \int_{t}  d^{3}r p^{0}e^{i\Delta p x}\frac{1}{(2\pi)^{3}} \int d^{4}u e^{-iux} \delta (pu) p^{0}a^{\dag}\left(\mathbf{p}-\frac{\mathbf{u}}{2}\right) a\left(\mathbf{p}+\frac{\mathbf{u}}{2}\right) = \nonumber \\
\int_{t} d^{4}u  p^{0} \delta^{(3)} (\Delta \textbf{p} - \textbf{u})e^{i(\Delta p_{0}-u_{0}) t}  \delta (pu) p^{0}a^{\dag}\left(\mathbf{p}-\frac{\mathbf{u}}{2}\right) a\left(\mathbf{p}+\frac{\mathbf{u}}{2}\right).
 \label{17}
\end{eqnarray}
Performing integration in the last expression, we finally obtain Eq. (\ref{15}).

Equations (\ref{11}), (\ref{12}) and  (\ref{16}) imply that,
in general, $f(x,p)$ depends
on $\lambda (x',p')$ in some nonlocal way. It is instructive  to  express $\lambda (x,p)$ through  $f(x,p)$ 
 by assuming that  $f(x,p)$ is sufficiently smooth 
in spatiotemporal coordinates across the freeze-out hypersurface. Then,  one can expect that
$\lambda (x',p')$ is also a smooth function.  Using the Taylor expansion around $x$, $\lambda (x',p') = \lambda (x,p') + ... $ and neglecting terms with derivatives, we get
\begin{eqnarray}
 \mathrm{Tr}[\hat{f}(x,p)\rho]\approx  \mathrm{Tr}[\hat{f}(x,p)\rho [x]],
 \label{18}
\end{eqnarray}
where 
\begin{eqnarray}
    \rho [x]=  \frac{1}{Z [x]}\exp \left (- \int d^{3}p'\frac{ p'^{\mu}}{p_{0}'}  \lambda (x,p') \int d\sigma'_{\mu}\hat{f}(x',p')\right ) ,
 \label{18.0}
\end{eqnarray}
and $Z [x]=\mathrm{Tr}\left[ \exp \left (- \int d^{3}p'\frac{ p'^{\mu}}{p_{0}'}  \lambda (x,p') \int d\sigma'_{\mu}\hat{f}(x',p')\right ) \right ]$.  Inserting (\ref{16}) in (\ref{18.0}) and using the Gauss theorem, we easily find 
\begin{align}
 &\int d^{3}p'\frac{ p'^{\mu}}{p_{0}'}  \lambda (x,p') \int_{\sigma} d\sigma'_{\mu}\frac{1}{(2\pi)^{3}} \int d^{4}u e^{-iux'} \delta (p'u) p^{0}a^{\dag}\left(\mathbf{p}'-\frac{\mathbf{u}}{2}\right) a\left(\mathbf{p}'+\frac{\mathbf{u}}{2}\right) =\nonumber \\ 
&\int d^{3}p'  \lambda (x,p') \int_{t} d^{3}r'\frac{1}{(2\pi)^{3}} \int d^{4}u e^{-iux'} \delta (p'u) p^{0}a^{\dag}\left(\mathbf{p}'-\frac{\mathbf{u}}{2}\right) a\left(\mathbf{p}'+\frac{\mathbf{u}}{2}\right) = \nonumber \\  
 &\int d^{3}p'\lambda (x,p')a^{\dag}(\mathbf{p}') a(\mathbf{p}'),  
 \label{18.1}
\end{align}
which gives 
\begin{eqnarray}
 \rho [x]= \frac{1}{Z [x]}\exp \left (- \int d^{3}p'  \lambda (x,p') a^{\dag}(\mathbf{p}') a(\mathbf{p}') \right ).
 \label{19.1}
\end{eqnarray}

Equations (\ref{11}), (\ref{18}), and (\ref{19.1}) allow us to express $\lambda (x,p)$ through  $f(x,p)$.  
We start by substituting  $\hat{f}(x,p)$, see Eq. (\ref{16}), into the right-hand side of Eq. (\ref{18}). Then, 
\begin{eqnarray}
 \mathrm{Tr}[\hat{f}(x,p)\rho [x]] = \nonumber \\
 \frac{1}{(2\pi)^{3}} \int d^{4}u e^{-iux} \delta (pu)  p^{0} \mathrm{Tr}\left[a^{\dag}\left(\mathbf{p}-\frac{\mathbf{u}}{2}\right) a\left(\mathbf{p}+\frac{\mathbf{u}}{2}\right) \rho [x] \right].
 \label{20}
\end{eqnarray}
Here,  $\mathrm{Tr}\left[a^{\dag}\left(\mathbf{p}-\frac{\mathbf{u}}{2}\right) a\left(\mathbf{p}+\frac{\mathbf{u}}{2}\right) \rho [x]  \right]$ can be calculated by means of  Gaudin's method \cite{Gaudin} (see also Ref. \cite{Groot}). Below, for the reader’s convenience, we
present an elementary derivation of it. 

 Let us start by defining 
 $a(\textbf{p},\alpha)$, 
\begin{eqnarray}
a(\textbf{p},\alpha) =  \exp \left ( \alpha \int d^{3}p'  \lambda (x,p') a^{\dag}(\mathbf{p}') a(\mathbf{p}')\right ) a(\textbf{p}) \exp \left ( - \alpha \int d^{3}p'  \lambda (x,p') a^{\dag}(\mathbf{p}') a(\mathbf{p}')\right ).
\label{21} 
\end{eqnarray}
Note that $a(\textbf{p}, 0) = a(\textbf{p}) $ and $\mathrm{Im}(\alpha) = 0$. 
Expression  (\ref{21}) implies that  $a(\textbf{p},\alpha)$ 
satisfies  equation 
\begin{eqnarray}
\frac{\partial a(\textbf{p},\alpha)}{\partial \alpha} = \left [     \int d^{3}p'  \lambda (x,p') a^{\dag}(\mathbf{p}') a(\mathbf{p}'), a(\textbf{p},\alpha) \right ].
\label{22} 
\end{eqnarray}
This  yields then 
\begin{eqnarray}
\frac{\partial a(\textbf{p},\alpha)}{\partial \alpha} =  -  \lambda (x,p) a(\textbf{p},\alpha).
\label{23} 
\end{eqnarray}
The solution of this equation is 
\begin{eqnarray}
 a(\textbf{p},\alpha) =  a(\textbf{p}) \exp{(- \alpha \lambda (x,p))}.
\label{24} 
\end{eqnarray}
 Using  the cyclic invariance of the trace
and Eqs. (\ref{19.1}) and (\ref{21}), we obtain 
\begin{eqnarray}
\mathrm{Tr} [\rho [x]  a^{\dag}(\textbf{p}_{1})a(\textbf{p}_{2}) ] = \mathrm{Tr} [a(\textbf{p}_{2}) \rho [x]  a^{\dag}(\textbf{p}_{1}) ] = \nonumber \\ \mathrm{Tr} [\rho [x] a(\textbf{p}_{2}, 1) a^{\dag}(\textbf{p}_{1}) ].
\label{25} 
\end{eqnarray}
Taking into account  Eqs. (\ref{3}) 
and  (\ref{24}),   the right-hand side  of the above equation can be rewritten as 
\begin{eqnarray}
 \mathrm{Tr} [\rho [x] a(\textbf{p}_{2}, 1) a^{\dag}(\textbf{p}_{1}) ]= \mathrm{Tr} [\rho [x]  a^{\dag}(\textbf{p}_{1})a(\textbf{p}_{2},1) ] + [a(\textbf{p}_{2},1),a^{\dag}(\textbf{p}_{1})] = \nonumber \\
  e^{- \lambda (x,p_{2})} \left( \mathrm{Tr} [\rho [x]  a^{\dag}(\textbf{p}_{1})a(\textbf{p}_{2}) ] + \delta^{(3)}(\textbf{p}_{2}-\textbf{p}_{1}) \right ).
\label{26} 
\end{eqnarray}
Substituting this into Eq. (\ref{25}) we find 
\begin{eqnarray}
\mathrm{Tr} [\rho [x]  a^{\dag}(\textbf{p}_{1})a(\textbf{p}_{2}) ]  = \delta^{(3)}(\textbf{p}_{1}-\textbf{p}_{2}) \frac{1}{e^{\lambda (x,(p_{1}+p_{2})/2)}-1}.
\label{27} 
\end{eqnarray}
Inserting (\ref{27}) in  (\ref{20}), we  obtain\footnote{An attentive reader can notice that $\langle \hat{f}(x,p)\rangle_{[x]}$  has the same form as 
for the ideal Bose gas, after substitution $\lambda (x,p) \rightarrow p_{\mu}\beta^{\mu}(x)$.}
\begin{eqnarray}
 \langle \hat{f}(x,p)\rangle_{[x]} = \frac{1}{(2\pi)^{3}}\frac{1}{e^{\lambda (x,p)}-1},
   \label{28}
\end{eqnarray}
where $\langle ...\rangle_{[x]}=\mathrm{Tr}[...\rho [x]]$. One observes that $\langle \hat{f}(x,p)\rangle_{[x]}$ is non-negative for $\lambda(x,p)\geq 0 $.
Taking into account Eqs. (\ref{11}) and (\ref{18}), we finally have 
\begin{eqnarray}
 \lambda (x,p) \approx \ln \left [ 1+ \frac{1}{(2\pi)^{3}f(x,p)} \right ] .
 \label{29}
\end{eqnarray}

\section{ One-particle momentum spectra and two-particle    correlations  in multiplicity-selected
collision events}
\label{HBT}
 
Without loss of generality, one can assume that the maximum entropy principle is applied at the post-freeze-out 
hypersurface  $t=\text{const}$. This implies that 
\begin{eqnarray}
\rho = \frac{1}{Z}\exp \left (- \int_{t} d^{3}r\int d^{3}p \lambda (x,p)\hat{f}(x,p)\right ).
 \label{30}
\end{eqnarray}
To explicitly calculate the particle momentum spectra, we replace the integral in Eq. (\ref{30})  by the sum as 
\begin{eqnarray}
\rho \approx \frac{1}{Z}\exp \left (- \sum_{s}\int d^{3}p \lambda (x_{s},p)\int_{v_{s}} d^{3}r\hat{f}(x,p)\right ).
\label{33} 
\end{eqnarray}
The integrals in the above equation are taken  over the homogeneity regions $v_{s}$ of the 
$\lambda(x,p) $  
around some points  $x_{s}^{\mu}=(t,\mathbf{r}_{s})$. The homogeneity region  is defined as
a  region  where $\lambda(x,p) $ 
does not vary noticeably. The key assumption underlying this  approximation 
is that the characteristic size $L$ of the corresponding volume element  
 is large enough from a microscopic point of view, meaning that the typical microscopic correlation
lengths are much smaller than the size of a cell, e.g., $L\gg 1/m$.
 
Inserting $\hat{f}(x,p)$, see Eq. (\ref{16}), into (\ref{33}) we find 
\begin{eqnarray}
\rho \approx \frac{1}{Z}\exp {\left (-\int d^{3}k d^{3}k'    A (\textbf{k}, \textbf{k}')a^{\dag}(\textbf{k})a(\textbf{k}') \right )}
\label{34} 
\end{eqnarray}
where $Z=\mathrm{Tr}[\exp {\left (-\int d^{3}k d^{3}k'    A (\textbf{k}, \textbf{k}')a^{\dag}(\textbf{k})a(\textbf{k}') \right )}]$, and 
\begin{eqnarray}
   A (\mathbf{k}, \mathbf{k}') =\sum_{s} A_{s} (\mathbf{k}, \mathbf{k}'), \label{34.1} \\
   A_{s} (\mathbf{k}, \mathbf{k}')=  \lambda(x_{s},(k+k')/2)  
\delta^{(3)}_{s}(\mathbf{k}-\mathbf{k}')
  ,\label{35.0} \nonumber \\
 \delta^{(3)}_{s}(\mathbf{k}-\mathbf{k}') = \frac{1}{(2\pi)^{3}}   \int_{v_{s}}d^{3}r  e^{i(k-k')x} \label{35}. 
\end{eqnarray}
Applying  a fixed
particle number constraint to the statistical operator $\rho$, we obtain the normalized statistical operator  conditional on $N$, 
\begin{eqnarray}
 \rho_{N} = \frac{{\cal P}_{N}\rho {\cal P}_{N}}{\mathrm{Tr}[{\cal P}_{N}\rho {\cal P}_{N}]}\approx \frac{1}{Z_{N}}{\cal P}_{N}\exp {\left (-\int d^{3}k d^{3}k'    A (\textbf{k}, \textbf{k}')a^{\dag}(\textbf{k})a(\textbf{k}') \right )} {\cal P}_{N},
   \label{35.1} 
\end{eqnarray}
where ${\cal P}_{N}$ is defined by Eq. (\ref{1}), and $Z_{N}=\mathrm{Tr}[{\cal P}_{N}\exp {\left (-\int d^{3}k d^{3}k'    A (\textbf{k}, \textbf{k}')a^{\dag}(\textbf{k})a(\textbf{k}') \right )} {\cal P}_{N}]$. One can see that the probability of producing $N$ particles, $p(N)$, is given by $p(N)=\mathrm{Tr}[{\cal P}_{N}\rho {\cal P}_{N}]=Z_{N}/Z$.
The calculations needed to compute the expectation values with the statistical operators  (\ref{34}) and (\ref{35.1}) are based on the assumption that $\delta^{(3)}_{s}(\mathbf{k}-\mathbf{k}')  $ is a sufficiently sharp function in the momentum difference for any $s$, and that 
\begin{eqnarray}
 \int d^{3}k \delta^{(3)}_{s}(\mathbf{k}_{1}-\mathbf{k})\delta^{(3)}_{s'}(\mathbf{k}-\mathbf{k}_{2})
 \approx \delta_{ss'}\delta^{(3)}_{s}(\mathbf{k}_{1}-\mathbf{k}_{2}),  \label{36}
\end{eqnarray}
where $\delta_{ss'}$ is  the Kronecker delta. 
Corresponding calculations can be done by adapting Gaudin's method \cite{Gaudin} 
to our problem. They are similar to those that were performed  in 
Ref. \cite{Akkelin-1}. Therefore, we only report the most important intermediate steps here.

We start by defining   $a(\textbf{p},\alpha)$, $a(\textbf{p}, 0) = a(\textbf{p}) $,  
 as 
\begin{eqnarray}
a(\textbf{p},\alpha) =  \exp {\left (\alpha \int d^{3}k d^{3}k'    A (\textbf{k}, \textbf{k}')a^{\dag}(\textbf{k})a(\textbf{k}') \right )} a(\textbf{p})\exp {\left (-\alpha \int d^{3}k d^{3}k'    A (\textbf{k}, \textbf{k}')a^{\dag}(\textbf{k})a(\textbf{k}') \right )}.
\label{37} 
\end{eqnarray}
Applying the operator identity 
\begin{eqnarray}
 e^{X}Ye^{-X} =Y + \left[ X, Y\right] + \frac{1}{2!}\left [X, \left[ X, Y \right]\right ] +\frac{1}{3!}\left[X, \left[X, \left[ X, Y\right]\right ] \right ] + ...,
\label{38} 
\end{eqnarray}
we can write the $a(\textbf{p},\alpha)$ as 
\begin{eqnarray}
     a(\textbf{p}, \alpha) = \int d^{3}k G_{\alpha}^{*}(\textbf{p}, \textbf{k}) a(\textbf{k}). \label{39} 
\end{eqnarray}
From Eqs. (\ref{37}) and (\ref{39}) we have that 
\begin{eqnarray}
\int d^{3}k G_{\alpha_{1}}^{*}(\textbf{p}_{2}, \textbf{k})G_{\alpha_{2}}^{*}(\textbf{k}, \textbf{p}_{1}) =  G_{\alpha_{1}+\alpha_{2}}^{*}(\textbf{p}_{2}, \textbf{p}_{1}).
\label{40} 
\end{eqnarray}
Taking into account (\ref{36}) and assuming that the size of a cell is much larger than the typical quantum correlation length,  we get 
\begin{eqnarray}
     G_{\alpha}^{*}(\textbf{p}, \textbf{k}) \approx
     \sum_{s}  \exp \left( -\alpha  \lambda(x_{s},(k+p)/2)  \right)  \delta^{(3)}_{s}(\textbf{p}-\textbf{k}). \label{41} 
\end{eqnarray}
One can see that 
\begin{eqnarray}
 G_{\alpha}^{*}(\textbf{p}, \textbf{k}) = G_{\alpha}(\textbf{k}, \textbf{p}).
\label{42} 
\end{eqnarray}
With the help of Eqs. (\ref{39}) and (\ref{41}), one can get expressions for the covariant one- and two-particle momentum spectra at fixed multiplicities.
They are given by 
\begin{eqnarray}
p_{0}\frac{d^{3}  N  }{d^{3}p} = p_{0}\langle a^{\dag}(\textbf{p}) a(\textbf{p}) \rangle_{N} 
\label{43}
\end{eqnarray}
and 
\begin{eqnarray}
p^{0}_{1}p^{0}_{2} \frac{d^{6}  N(N-1) }{d^{3}p_{1}d^{3}p_{2}}=p^{0}_{1}p^{0}_{2} \langle a^{\dag}(\textbf{p}_{1})a^{\dag}(\textbf{p}_{2}) a(\textbf{p}_{1})a(\textbf{p}_{2}) \rangle_{N},
\label{44}
\end{eqnarray}
respectively, and we define  $\langle ...\rangle_{N}=\mathrm{Tr}[...\rho_{N}]$.
Using the cyclic property of a trace and  Eqs. (\ref{37}), (\ref{39}), and (\ref{41}) for $\alpha=1$, one can perform calculations of the corresponding expectation values. It results  in 
\begin{eqnarray}
\langle a^{\dag}(\textbf{p}_{1}) a(\textbf{p}_{2}) \rangle_{N} = \sum_{n=1}^{N}\frac{Z_{N-n}}{Z_{N}}  G_{n}^{*}(\textbf{p}_{2}, \textbf{p}_{1}),
 \label{45}
\end{eqnarray}
and 
\begin{eqnarray}
\langle a^{\dag}(\textbf{p}_{1})a^{\dag}(\textbf{p}_{2}) a(\textbf{p}_{1})a(\textbf{p}_{2}) \rangle_{N}= \nonumber \\
 \sum_{n=1}^{N-1} \sum_{s=1}^{N-n}  \frac{Z_{N-n-s}}{Z_{N}} \left( G_{n}^{*}(\textbf{p}_{2}, \textbf{p}_{2}) G_{s}^{*}(\textbf{p}_{1}, \textbf{p}_{1}) + G_{n}^{*}(\textbf{p}_{2}, \textbf{p}_{1}) G_{s}^{*}(\textbf{p}_{1}, \textbf{p}_{2}) \right).
 \label{46}
\end{eqnarray}
Here $G_{\alpha} (\textbf{p}_{1}, \textbf{p}_{2})$ is defined by Eq. (\ref{41}).
Replacing sums over cells with the integral, we  get 
\begin{eqnarray}
G_{\alpha} (\textbf{p}_{1}, \textbf{p}_{2}) \approx  \frac{1}{(2\pi)^{3}} \int_{t}  d^{3}r e^{-i(p_{1}-p_{2})x} e^{-n\lambda (x,p)} ,
\label{47}
\end{eqnarray}
where $p^{\mu}=(p^{\mu}_{1}+p^{\mu}_{2})/2$.
To evaluate  Eqs. (\ref{45}) and (\ref{46}), we  need explicit  expressions for
 $Z_{n}$. They can be evaluated as follows. First, note that the definition of $\rho_{N}$
means that 
\begin{eqnarray}
\int d^{3}p \langle a^{\dag}(\textbf{p}) a(\textbf{p}) \rangle_{N} = N.
 \label{48}
\end{eqnarray}
Then, accounting for Eq. (\ref{45}), we get the recursive formula 
\begin{eqnarray}
nZ_{n} =  \sum_{s=1}^{n}Z_{n-s}\int d^{3}p G_{s}^{*}(\textbf{p}, \textbf{p}),
\label{49} 
\end{eqnarray}
where $Z_{0}=1$ by definition. 

Nontrivial information about the spatiotemporal structure of particle emission is encoded in the 
two-particle correlation function (see e.g. Refs. \cite{HBT-1,HBT-2,HBT-3}), which  is defined as the ratio of the two-particle momentum
spectrum to one-particle ones
and can be evaluated as
\begin{eqnarray}
C_{N}(\textbf{p}_{1},\textbf{p}_{2}) =\frac{ \langle a^{\dag}(\textbf{p}_{1})a^{\dag}(\textbf{p}_{2}) a(\textbf{p}_{1})a(\textbf{p}_{2}) \rangle_{N}}{\langle a^{\dag}(\textbf{p}_{1})a(\textbf{p}_{1})  \rangle_{N} \langle a^{\dag}(\textbf{p}_{2})a(\textbf{p}_{2})  \rangle_{N}  }, 
\label{50} 
\end{eqnarray}
where corresponding expectation values are defined in Eqs. 
(\ref{45}) and  (\ref{46}). 

For completeness, we also present here expressions for the average   one-particle
\begin{eqnarray}
p_{0}\frac{d^{3} \langle N \rangle }{d^{3}p} = p_{0}\langle a^{\dag}(\textbf{p}) a(\textbf{p}) \rangle 
\label{51}
\end{eqnarray}
and two-particle 
\begin{eqnarray}
p^{0}_{1}p^{0}_{2} \frac{d^{6} \langle N(N-1)\rangle }{d^{3}p_{1}d^{3}p_{2}}=p^{0}_{1}p^{0}_{2} \langle a^{\dag}(\textbf{p}_{1})a^{\dag}(\textbf{p}_{2}) a(\textbf{p}_{1})a(\textbf{p}_{2}) \rangle
\label{52}
\end{eqnarray}
momentum
spectra, corresponding to some
fixed centrality of a collision (in other words, corresponding to collisions with some fixed
initial size of the reaction zone).
Using $\frac{d^{3}\langle N \rangle}{d^{3}p} =    \int_{t}  d^{3} r f(x,p)$ and $f(x,p) = \langle \hat{f}(x,p) \rangle \approx \langle \hat{f}(x,p)\rangle_{[x]} $, where $ \langle \hat{f}(x,p)\rangle_{[x]} $ is defined in Eq. (\ref{28}), we  get 
\begin{eqnarray}
\frac{d^{3}\langle N \rangle}{d^{3}p}  \approx \frac{1}{(2\pi)^{3}}  \int_{t}  d^{3}r\frac{1}{e^{\lambda (x,p)}-1},
\label{53} 
\end{eqnarray}
Next, we calculate the average two-particle momentum spectra, see Eq. (\ref{52}).
One can show (see, e.g., Ref. \cite{Akkelin-1}) that, for the statistical operator in the form of Eq. (\ref{34}),  
Wick's \cite{Wick} decomposition
\begin{eqnarray}
\langle a^{\dag}(\textbf{p}_{1})a^{\dag}(\textbf{p}_{2}) a(\textbf{p}_{1})a(\textbf{p}_{2}) \rangle=
\langle a^{\dag}(\textbf{p}_{1})a(\textbf{p}_{1})\rangle \langle a^{\dag}(\textbf{p}_{2})a(\textbf{p}_{2}) \rangle +
\langle a^{\dag}(\textbf{p}_{1})a(\textbf{p}_{2})\rangle \langle a^{\dag}(\textbf{p}_{2})a(\textbf{p}_{1}) \rangle 
\label{54} 
\end{eqnarray}
can be applied. 
Using (\ref{15}), we get 
\begin{eqnarray}
\langle a^{\dag}(\textbf{p}_{1}) a(\textbf{p}_{2}) \rangle  \approx   \frac{1}{(2\pi)^{3}} \int_{t}  d^{3}r e^{i(p_{2}-p_{1})x}\frac{1}{e^{\lambda (x,p)}-1}.
\label{55} 
\end{eqnarray}
By substituting (\ref{55}) into Eq. (\ref{54}), we get the expression for the average two-particle momentum spectra and,
thereby, for the corresponding correlation function 
\begin{eqnarray}
C(\textbf{p}_{1},\textbf{p}_{2}) =\frac{p^{0}_{1}p^{0}_{2} \frac{d^{6} \langle N(N-1) \rangle }{d^{3}p_{1}d^{3}p_{2}}}{p^{0}_{1}\frac{d^{3} \langle  N  \rangle  }{d^{3}p_{1}} p^{0}_{2}\frac{d^{3} \langle  N \rangle }{d^{3}p_{2}} } =\frac{ \langle a^{\dag}(\textbf{p}_{1})a^{\dag}(\textbf{p}_{2}) a(\textbf{p}_{1})a(\textbf{p}_{2}) \rangle}{\langle a^{\dag}(\textbf{p}_{1})a(\textbf{p}_{1})  \rangle \langle a^{\dag}(\textbf{p}_{2})a(\textbf{p}_{2})  \rangle  }.
\label{56} 
\end{eqnarray}

The width of the correlation functions in $|\mathbf{p}_{1} - \mathbf{p}_{2}|$
reflects the space-time structure of the underlying source at a
fixed $\textbf{p}_{1} + \textbf{p}_{2}$. This structure can be inferred by suitably parametrizing
the correlation function. One can notice, see $C_{N}(\textbf{p}_{1},\textbf{p}_{2})$ and $C(\textbf{p}_{1},\textbf{p}_{2})$,  that spatiotemporal information about the source is encoded in the function $\lambda (x,p)$. The latter is related to the one-particle  Wigner distribution $f(x,p)$, see Eq. (\ref{29}).  This implies that $\lambda (x,p)$ depends on $\langle N \rangle$ and does 
not depend on $N$. A notable feature of Eq. (\ref{50}) is that $N$ does not contribute to the effective size of the system: this size increases only with  $\langle N \rangle$, and reaches its maximal value for 
the central collisions with maximal geometric size of the initial reaction zone. As a result, one can conclude that 
multiplicity dependence of the correlation radii (fit parameters to the 
correlation function) for high  $N \gg \langle N \rangle$, where $\langle N \rangle$ corresponds to the collisions where optimal overlapping of colliding protons is reached, should be weak.
 This effect should be particularly relevant for the interpretation of the 
independence of correlation 
radius parameters on the charged-particle multiplicity in high-multiplicity $pp$ collisions at a fixed energy of the LHC \cite{Atlas,CMS}.

\section{ Conclusions }
\label{Concl}

In this paper, we argue that multiplicity fluctuations of quantum origin dominate the fluctuations of the particle multiplicity in high-multiplicity collision events.    Quantum-to-classical transition of multiplicity fluctuations happens when the interference term between the corresponding states becomes negligible. Until this happens, superposition cannot be interpreted as a classical ensemble of its component states. We introduce the statistical ensemble of the multiplicity fluctuations of particles in
high-multiplicity events after the system's freeze-out by applying the maximum entropy 
kinetic freeze-out prescription. We study how multiplicity fluctuations at high multiplicities impact the multiplicity-selected particle momentum spectra and correlation functions. In particular, we find that the effective system’s volume, when extracted from the fit of the correlation function,  should become approximately independent of the
particle multiplicity in the limit of very high multiplicities in multiplicity-selected collision events.
The results of our analysis  should be taken into account when theoretical models are 
compared with multiplicity-selected particle momentum spectra and correlations in high-multiplicity $pp$ collision events at a fixed energy of the LHC.

\begin{acknowledgments}
 S.V.A. gratefully
acknowledges support from the Simons Foundation (Grant
No. SFI-PD-Ukraine-00014578).
\end{acknowledgments}

\end{document}